# NFL: Robust Learned Index via Distribution Transformation

Shangyu Wu
City University of Hong Kong
Hong Kong
shangyuwu2-c@my.cityu.edu.hk

Yufei Cui*
City University of Hong Kong
Hong Kong
yufeicui3-c@my.cityu.edu.hk

Jinghuan Yu
City University of Hong Kong
Hong Kong
jinghuayu2-c@my.cityu.edu.hk

Xuan Sun
City University of Hong Kong
Hong Kong
xuansun-c@my.cityu.edu.hk

Tei-Wei Kuo
City University of Hong Kong
Hong Kong
teiwei.kuo@cityu.edu.hk

Chun Jason Xue
City University of Hong Kong
Hong Kong
jasonxue@cityu.edu.hk

## ABSTRACT

Recent works on learned index open a new direction for the indexing field. The key insight of the learned index is to approximate the mapping between keys and positions with piece-wise linear functions. Such methods require partitioning key space for a better approximation. Although lots of heuristics are proposed to improve the approximation quality, the bottleneck is that the segmentation overheads could hinder the overall performance.

This paper tackles the approximation problem by applying a *distribution transformation* to the keys before constructing the learned index. A two-stage Normalizing-Flow-based Learned index framework (NFL) is proposed, which first transforms the original complex key distribution into a near-uniform distribution, then builds a learned index leveraging the transformed keys. For effective distribution transformation, we propose a Numerical Normalizing Flow (Numerical NF). Based on the characteristics of the transformed keys, we propose a robust After-Flow Learned Index (AFLI). To validate the performance, comprehensive evaluations are conducted on both synthetic and real-world workloads, which shows that the proposed NFL produces the highest throughput and the lowest tail latency compared to the state-of-the-art learned indexes.

## 1 INTRODUCTION

Learned Index [22], as an advance replacement of conventional indexes, has attracted a lot of attention in recent years. It leverages machine learning models to learn the mapping between keys and corresponding positions, greatly accelerating the data retrieval process. Existing learned indexes [6, 12, 13, 42] further address the updatable issue in the original learned index, and achieve significant improvements on various workloads. The key insight of these learned indexes is to regard linear models as a piece-wise linear function to approximate the cumulative distribution function (CDF). To obtain a good approximation, existing learned indexes propose various segmentation methods, e.g., the convex-hull-based segmentation [12]. However, such heuristic segmentation not only introduces extra time and space overhead, but also requires lots of efforts to design for both the algorithm and supportive operations. This paper solve the approximation problem from a different angle.

---

*Corresponding author.
Accepted as a conference paper in VLDB 2022.

In this work, we propose a new methodology which transforms the original keys to a near-uniform key space before constructing the learned index, so that the learned index can make a much better approximation on the CDF. This distribution transformation tackles the approximation problem from the root, and significantly improve the performance for all aspects of learned indexes. Normalizing flows (NFs) [8, 34], as a family of generative models, could be applied to perform *distribution transformation.*

To reach the goal of distribution transformation, we propose a two-stage Normalizing-Flow-Learned index framework (NFL), consisting of a Numerical Normalizing Flow (Numerical NF) for distribution transformation and a robust After-Flow Learned Index (AFLI) for CDF approximation. Instead of directly segmenting the CDF curve, the proposed NFL first leverages the Numerical NF to transform the original keys into the near-uniform distributed keys, so that the CDF curve becomes roughly linear. Then, based on the transformed keys, the proposed AFLI could make a decent approximation on the transformed CDF. In addition, we also propose a new type of conflict degree metric to evaluate the transformation quality of Numerical NF.

Existing normalizing flows are not applicable for the case of learned indexes, as they are designed for high dimensional data with rich semantic or spatial features, e.g., computer vision [1, 41]. However, the learned indexes deal with numerical data with less complexity for learning. Furthermore, existing normalizing flows are of too high complexity (e.g., 4 layers, 16 parameters) for processing the queries. In the proposed Numerical NF, for the efficacy of NFs, we propose a feature space expansion scheme for enriching the learnable features. Moreover, a set of optimizations for efficiency is proposed, including a switching mechanism based on *conflict degree.*

After the transformation of Numerical NF, considering the characteristics of the transformed keys, the proposed AFLI only needs a simple and efficient structure to handle local conflicts, providing a robust performance in both throughput and tail latency. In AFLI, we replaces the complex and expensive adjustments in existing learned indexes with simple *Modelling* operations, which turns a dense array into a model-based node.

Finally, we conduct a series of experiments on both synthetic and real-world workloads. The proposed NFL and learned index are compared with the state-of-the-art learned indexes in terms of throughput, tail latency, index size, and bulk loading time. Experimental results show that the proposed NFL beats all existing learned index in almost all kinds of workloads with the highest throughput and the lowest tail latency. Specifically, the proposed NFL can achieve 2.77x improvements on average in throughput and 43% reductions on average in tail latency.

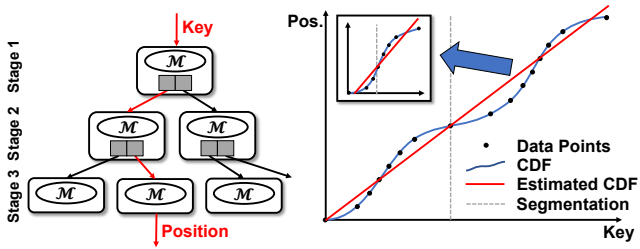

Figure 1: The structure and the insight of the recursive model index.

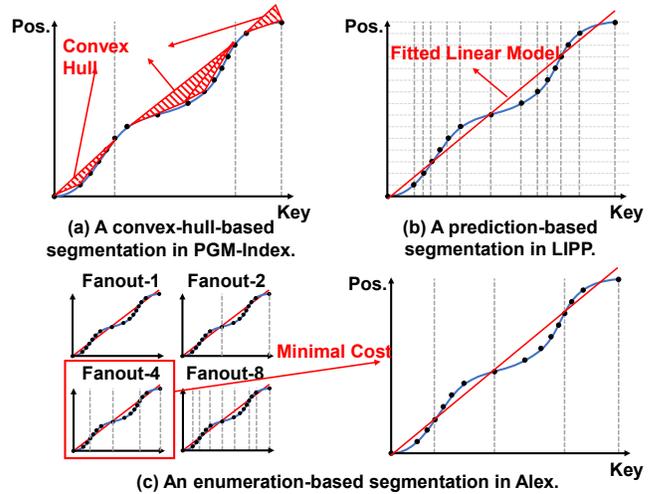

Figure 2: Different segmentation on the CDF.

The main contributions of this paper are:

- To the best of our knowledge, this is the first work that explores the *distribution transformation* to address the approximation challenge of the learned index.
- We propose a two-stage normalizing-flow-based learned index (NFL) framework, which transforms keys into another key space, where the learned index could be easier to fit a more linear CDF curve.
- To improve the transformation quality of NF in the NFL, we propose a numerical normalizing flow, which could enrich the poor feature in the indexing problem. We also make an efficient implementation of NF.
- After the distribution transformation of NF, based on the characteristics of the transformed keys, we propose an After-Flow Learned Index (AFLI), which can efficiently support almost all indexing operations, providing a robust performance in both throughput and tail latency.
- We conduct a set of experiments with both synthetic and real-world workloads to show the effectiveness and the efficiency of the proposed NFL framework.

The rest of this paper is organized as follows. Section 2 introduces the background and motivations. Section 3 presents the proposed normalizing-flow-learned index framework. Specifically, Section 3.2 introduces the numerical normalizing flow in the NFL for distribution transformation. Section 3.3 presents the after-flow learned index in detail that makes a better approximation on the transformed distribution. Section 4 presents the experimental results with discussion. Section 5 presents the literature review. Finally, in Section 6, we conclude this paper and discuss the future work.

## 2 BACKGROUND AND MOTIVATIONS
### 2.1 Preliminary
Recent works on Learned Index (LI) [22] observed that the mapping between keys and the corresponding positions can be formulated as a Cumulative Distribution Function (CDF) model,

$$\text{pos} = F(\text{key}) * N \qquad (1)$$

where pos is the predicted position, $F(\cdot)$ is the approximated CDF that estimates the likelihood $p(x \leq \text{key})$ of a key $x$ smaller than or equal to the given key, and $N$ is the total number of keys. Based on the formulation, learned index can leverage ML models to approximate the CDF to predict the position, replacing multiple comparison operations in BTree with computations.

Since using a single model (e.g., a fully connected neural network) to approximate the CDF suffers the "last mile" issue, i.e., much more space and time are required to reduce the error further from thousands to hundreds, therefore, Kraska et al. [22] propose the *Recursive Model Index (RMI)*, which is a hierarchy of linear models. As shown in the left of Figure 1, the higher-level model takes the key as input and picks the model at the next level, and so on. Later, the leaf model predicts the estimated position of the input key. Due to prediction errors, the RMI will finally use local search (e.g., binary search) to correct the estimated position.

From the description of the search process, we can find that the approximation is actually constructing a better piece-wise linear function, reducing amount of local search. As shown in the right of Figure 1, the root model in stage 1 first tries to approximate the curve of the CDF. But, since using two linear models, jointly acting as a piece-wise linear function, can make a better approximation of the curve, the root model segments the curve of the CDF into two sub-curves. The first sub-curve is then approximated by the first child model in the stage 2. Similarly, the first sub-curve is also segmented into two sub-curves, and so do other sub-curves. Compared to higher-level models, next-level models make a better approximation with less prediction errors as more segments are used to approximate the curve, so that the time cost of local search can be reduced. However, a deeper hierarchy/tree prolongs the traversal time, as it increases the number of stages thus the number of predictions of the ML model.

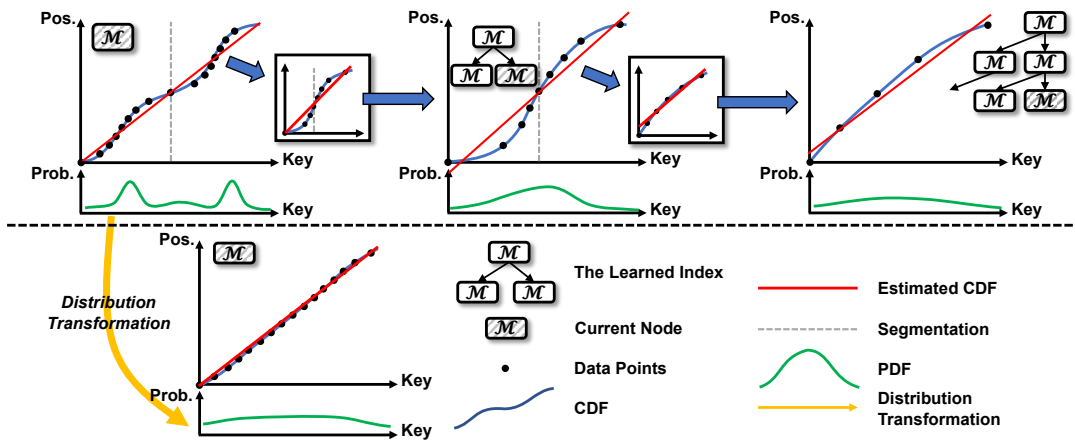

Figure 3: Applying a transformation on the CDF.

Inspired by the RMI, the follow-up learned indexes [6, 12, 42] not only address the updatable issues in the RMI, but also improve the approximation quality of linear models, making the index deliver a comparable performance on various indexing operations. As shown in the Figure 2(a), PGM-Index [12] propose a convex-hull-based segmentation to partition the key space. Specifically, the key points on the curve can be aggregated by a convex hull. If the convex hull cannot be enclosed in a rectangle of height no more than $2\epsilon$, the set of keys in the convex hull are allocated to a segment. Then a new convex hull is created for the subsequent segmentation. Although the segmentation in PGM-Index can ensure that the prediction error is less than $\epsilon$, the approximation quality is limited as the segmentation doesn't refer to the global information of the CDF. Moreover, how to choose the hyper-parameter $\epsilon$ is also a key point.

Different from PGM-Index, ALEX [6] and LIPP [42] first fit a linear model to learn the outline of the CDF, then segment the curve of the CDF based on predictions. Compared to prediction errors, LIPP is more concerned with *conflict degrees*, i.e., the number of keys predicted to the same position. As shown in the Figure 2(b), LIPP allocates all conflicted keys to the same segment. Figure 2(c) shows the main idea of segmentation in ALEX. ALEX exhaustively searches for the best number of segments. Specifically, before building the index, it tries different number of segments, namely fanout, calculates the cost of each kind of fanout, and chooses the best fanout with the minimal cost. Both segmentation in ALEX and LIPP significantly improve the approximation quality so that deliver a high performance.

All these variants of learned index pay their attention on how to partition the key space by designing better heuristics, to obtain better piece-wise linear approximations for the CDF. However, such segmentation requires a long period of time to process and large pre-allocated space to maintain. Furthermore, each learned index requires designing a set of supportive operations and optimizing their hyper-parameters by tuning.

## 2.2 Motivation

In this work, we explore in direction which is different from previous works. Instead of directly approximating the original CDF with a piece-wise linear function, the proposed method aims at transforming the original keys to another key space, where the learned index could work on a less complex key distribution. The proposed method is adaptive to different workloads, thus it alleviates the heavy dependence of expertise for designing heuristics and tuning hyper-parameters.

Conceptually, previous works segment the curve of the CDF into several sub-curves, in order to make every linear model fit the sub-curve better, so that the learned index can make a good approximation with a piece-wise linear function on the whole curve (Figure 3). The insight of such methods is to find a roughly linear sub-curve by repeatedly segmenting the curve, whose corresponding probability density function (PDF) is a local near-uniform distribution. To reach this goal, the learned index has to construct a deeper hierarchy, resulting in more traversal time and number of predictions. However, if there is an ideal key space where keys already satisfy a global near-uniform distribution, whose corresponding CDF curve is roughly linear, a single linear model is sufficient to fit such curve. Therefore, this motivates us to transform the original key space to a near-uniform key space by *distribution transformation*, before constructing the learned index.

Table 1: Statistics of ALEX [6] on longlat (LLT) and facebook (FB) workloads. Throughput are presented in million ops/sec.

|     |                     | Without NF  | With NF     |
| --- | ------------------- | ----------- | ----------- |
|     | Max Tree Height     | 4           | 3           |
|     | Average Tree Height | 2.30        | 2.01        |
| LLT | # Prediction Errors | 925,487,063 | 118,833,075 |
|     | # Predictions       | 124,831,692 | 101,108,381 |
|     | Throughput          | 8.21        | 11.57       |
|     | Max Tree Height     | 11          | 3           |
|     | Average Tree Height | 5.91        | 2.02        |
| FB  | # Prediction Errors | 928,113,206 | 453,003,864 |
|     | # Predictions       | 492,112,591 | 100,462,387 |
|     | Throughput          | 4.13        | 9.16        |

In practice, there is no unified heuristic that works for transforming heterogeneous key space in real-world workloads. Based on the *Change of Variables Theorem*, normalizing flow [34] can learn the transformation between a base distribution and any given data distribution. Previous NFs are only designed for high dimensional data with rich semantic or spatial features, e.g., computer vision [1, 41]. For learned indexes, the keys are numerical data and with less useful features. We propose the first numerical NF that transforms the keys into a global near-uniform distribution. We illustrate the efficacy of the proposed numerical NF on real-world workloads and a representative learned index. Table 1 shows the statistics of ALEX on two workloads with a highly non-linear CDF curve. After applying the proposed NF to transform the keys, the tree height of ALEX can be reduced from 11 to 3 levels on the facebook workload. On longlat workload, although the reduction of tree levels is $4 - 3 = 1$, the prediction errors are greatly reduced as NF generates more uniform distributed keys which reduces the number of out-of-boundary keys within each node. Besides, the numbers of predictions are also reduced. Therefore, ALEX can achieves great improvements on the throughput. The above results and analysis jointly demonstrate the importance of distribution transformation.

## 2.3 Normalizing Flows

Normalizing Flows [34] are a family of generative models, which transform a latent distribution $p(z)$ to a new distribution $p_G(x)$ by a series of parameterized generators, where $z$ and $x$ are corresponding random variables (r.v.). In our case, $z$ and $x$ corresponds to the ideal (transformed) and original keys. We define $\{z_i\}_i^M$ and $\{x_i\}_i^M$ as the samples of the two r.v., where $M$ is the size of the sample sets. We want the generated distribution $p_G(x)$ to approximate the given data distribution $p_{data}(x)$. For training the NF, the objective is to maximize the likelihood of the generated distribution $p_G(x)$ or optimize the Kullback-Leibler (KL) divergence between the generated distribution $p_G(x)$ and the data distribution $p_{data}(x)$,

$$G^* = \text{argmax}_G \mathbb{E}_{x \sim p_{data}} \log p_G(x) \quad (2)$$
$$G^* = \text{argmin}_G \mathbb{KL}(p_G \| p_{data}) \quad (3)$$

where $G = G_\theta = G_\theta(\cdot)$ is the parameterized generator, which takes a sample of $z$ and push-forward it to a sample of $x$, $x_i = G_\theta(z_i)$. $\theta$ represents the parameters. $G^*$ is the optimal generator.

The key advantage of NFs over other generative models (e.g., variational autoencoder [19]) is that, the generator $G$ is an invertible function. Once we have a well-trained $G$, $G^{-1}$ could be obtained easily by taking the inverse. Then, $G^{-1}$ could be used to encode the original key $x_i$ to the ideal key $z_i$, which conforms with our idea of *distribution transformation*.

Specifically, since $G_\theta$ is invertible, using the *Change of Variables Theorem*, we can derive,

$$p_G(x) = \left| \frac{\partial G_\theta(z)}{\partial z} \right|^{-1} p(z) \quad (4)$$

where $\left| \frac{\partial G_\theta(z)}{\partial z} \right|$ is the determinant of $G_\theta$'s Jacobian matrix. Another important point is that the determinant of the Jacobian matrix must be enough cheap to compute, otherwise the NF might introduce non-negligible overhead.

Previous NFs [9, 17] present various inveritble generators with efficient computations of the determinants. Considering a high-dimensional random variable, Kingma et al. [18], Papamakarios et al. [32] combine autoregressive models and NFs to learn the hidden states from previous dimensions and improve the transformation. To make invertible functions more expressive, Cao et al. [2], Huang et al. [16] propose to learn a more complex bijection using a monotonic neural network and achieve the state-of-the-art performance on various datasets.

## 2.4 Challenges

Motivated by distribution transformation, this work targets at providing a novel two-stage framework, called Normalizing-Flow-Learned Index (NFL), that first transforms the key distribution by normalizing flow, then build an after-flow learned index that effectively leverages the transformed key distribution. However, this process is not straightforward and we list the following challenges:

**Efficacy of normalizing flow:** Naively using NF is limited in a few ways: 1) the NF perform poorly due to limited features from the numerical data of keys; 2) the uniform distribution is hard to function directly as an training objective. We design a Numerical Normalizing Flow (Numerical NF) with a enriched feature space (see Section 3.2.1) and an easy-to-operate training objective.

**Efficiency of normalizing flow:** The transformation must be an efficient online step. Such requirement also limits the complexity of normalizing flows. Directly reducing the number of parameters in normalizing flows might degrade the transformation quality so that learned indexes require deeper hierarchy and more models to approximate the CDF. We design a set of optimizations for efficiency, without losing the efficacy of the NF. 3.2.2).

**Lack of proper indexes for transformed keys:** With the transformation of Numerical NF that fundamentally makes linear models approximate better, the design of learned indexes should be reconsidered in a new perspective. Therefore, based on the characteristics of the transformed data distribution, the locality of the transformed data distribution should be considered in the design of the learned index. We propose a After-Flow Learned Index (AFLI) for fully leveraging the transformed keys. (see Section pay3.3).

## 3 NFL: THE NORMALIZING-FLOW-LEARNED INDEX FRAMEWORK

### 3.1 Framework

Figure 4 shows the structure and the workflow of the Normalizing-Flow-Learned Index framework (NFL). The framework consists of two parts, a normalizing flow for distribution transformation and a learned index for CDF approximation. The input keys are first fed into the normalizing flow which transforms them to a near-uniform distribution. Then all transformed keys are used for building linear models in the learned index. Since batching requests (e.g, batching queries, batching insertions) is a common case in modern database [14, 15, 28, 29, 33], our NFL also processes requests in batches.

The theoretical derivation is as follows. Given an input key $x$ satisfying a certain distribution $p(x)$, the normalizing flow first transform it to a target distribution $p(z)$, i.e., $z = G_\theta(x)$. Note that we don't need to know the actual form of the distribution $p(x)$ as the normalizing flow only needs to maximize the likelihood of the transformed data on the target distribution. Then, the learned index takes the transformed key $z$ as the input, and predicts its position, i.e., $pos = F(z) * N$.

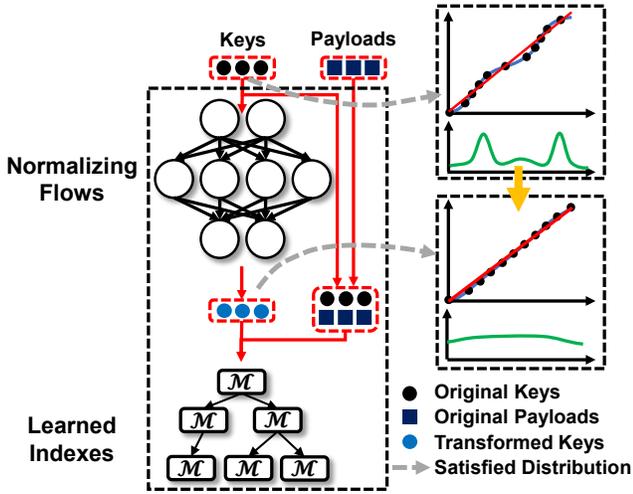

**Figure 4: The structure and the workflow of the NFL.**

The NFL offers a new perspective to fundamentally improve the learned indexes.

- The NFL is a unified framework universally applicable to any workload, due to the transformation capability and adaptability of flow-based generative models. It amortizes the requirement of domain expertise of designing piecewise approximation heuristics by splitting it into two simpler stages.
- The NFL is measurable and flexible. One implementation of NFL can be measured by the transformation quality of the normalizing flow. A consistently high-quality transformation makes an easier implementation of learned index.

Under this framework, we propose the generic prototypes of normalizing flow (Numerical NF) and learned indexes (AFLI). Before going to the details of specific designs, we present a quantitative evaluation metric for the transformation quality.

*3.1.1 Conflict Degree.* Since the log probability of the transformed distribution in the NF cannot accurately evaluate how nearly uniform the distribution should be for the learned index to deliver a high performance, we need a new metric that connects the transformation quality with the performance of the learned index. Considering that placing data in the predicted positions can eliminate prediction errors, we are motivated to introduce the conflict degree and a new metric called tail conflict degree to jointly quantify how nearly uniform the distribution is and how well the learned index would perform.

First, we introduce the conflict degree on each position.

DEFINITION 3.1. *For a set of data $X = \{x_1, \ldots, x_n\}$ and a fitted linear model $\mathcal{M}$, the conflict degree of position $j$ is:*

$$D_{\mathcal{M}}^{j} = |\{x_i \in X | \mathcal{M}(x_i) == j\}| \qquad (5)$$

*where $j$ ranges from $MIN(\{\mathcal{M}(x_i)\})$ to $MAX(\{\mathcal{M}(x_i)\})$.*

With the definition of the conflict degree, we further define the tail conflict degree to evaluate the transformation quality. The tail conflict degree indicates the upper bound of the conflicts for most positions.

DEFINITION 3.2. *For $m$ positions whose conflict degree is greater than 0, we represent their conflict degree with $\{D_{\mathcal{M}}^{j}\}$. Then, for a given tail percent $\gamma$, we let $t = INT(m \times \gamma)$, where INT represents the flooring operation. The **Tail Conflict Degree** based on a tail percent $\gamma$ is t-th larger conflict degrees among $\{D_{\mathcal{M}}^{j}\}$, represented by $D_{\mathcal{M}}^{\gamma}$.*

In this paper, we set the tail percent to 0.99. For example, given 1000 positions $\{j_1, \ldots, j_{1000}\}$ with conflict degrees $\{D_{\mathcal{M}}^{j_1}, \ldots, D_{\mathcal{M}}^{j_{1000}}\}$, based on the tail percent $\gamma = 0.99$, $t = INT(1000 \times 0.99) = 990$. The tail conflict degree is 990-th larger conflict degree. The tail conflict degree is a soft measurement. It could be useful in: 1) determine the execution of flow (see Section 3.2.2); 2) determine the capacity threshold of a node (see Section 3.3.1).

### 3.2 Numerical Normalizing Flow

The proposed Numerical NF aims to transform the key distribution into a near-uniform distribution. Since training normalizing flow with a uniform distribution as the objective might encounter the "Nan-loss" issue or the "INF-loss" issue, we replace it with a near-uniform distribution, i.e., a normal distribution with a large variance. After that, we then focus on feature engineering for efficacy (See Section 3.2.1), model slimming and process optimization for efficiency (See Section 3.2.2).

*3.2.1 Feature Space Expansion.* Existing NFs are used in computer vision or natural language process to process high-dimensional images or texts. Those kinds of data can provide rich features, e.g., semantic features or spatial features in images and texts. For example, in image generation tasks like hand-written digits, an NF generating number "9" could capture the spatial correlation among different pixels, and the semantic knowledge. The performance could be improved by feeding more images of number "9" for training. However, the data of keys are numerical data with hardly any useful knowledge. As the keys are required to be unique, feeding more data would increase the burden of an NF rather than help training. Therefore, one major target of our design is to enrich the features that the an NF could learn from, while maintaining high efficiency.

Algorithm 3.1 shows the whole process of *key transformation*. First, all input keys are normalized with a scaled min-max normalization to avoid keys without integral or floating part (Line 2). After the normalization, the algorithm start to expand feature space by repeatedly obtaining the integral part and the final floating part of the normalized keys (Line 3-17). After the feature expansion, the high-dimensional keys are fed into the NF (Line 18). Finally, the decoding layer merges the high-dimensional features into 1D keys (Line 19-22) and passes them to the learned index (Line 23). The time complexity of the feature expansion is $O(n \times d)$. Such brute-force key space expansion provides a 1-to-1 mapping from 1-D keys to d-dimensional keys, while maintaining a linear complexity.

*3.2.2 Efficient Processing of Normalizing Flows.* Different from linear models, normalizing flows do not have an analytical solution

**Algorithm 3.1** Key Distribution Transformation.

**Input:** A set of sorted keys $X = \{x_1, \ldots, x_n\}$ for bulk loading, the target dimension $d$, the scale factor $\theta$, the normalizing flow $F$.
**Output:** A set of transformed keys $Z = \{z_1, \ldots, z_n\}$.

1: $X^d = \emptyset$;
2: $\mu = \min(X), \sigma = \frac{\max(X) - \min(X)}{\theta}$;
3: **for** $x_i \in X$ **do**
4:    $x_{\text{norm}} = \frac{x_i - \mu}{\sigma}$;  /* Encoder, expand features for NF */
5:    $x_i^{\text{vec}} = [\,]$;
6:    $x_{\text{int}} = \text{INT}(x_i)$;
7:    $x_{\text{float}} = x_i - \text{INT}(x_i)$;
8:    Add $x_{\text{int}}$ into $x_i^{\text{vec}}$;
9:    **for** $k$ from 1 to $d - 2$ **do**
10:      Add $x_{\text{int}}$ into $x_i^{\text{vec}}$;
11:      $x_{\text{float}} = x_{\text{float}} * \theta$
12:      $x_{\text{int}} = \text{INT}(x_{\text{float}})$;
13:      $x_{\text{float}} = x_{\text{float}} - x_{\text{int}}$;
14:    **end for**
15:    Add $x_{\text{float}}$ into $x_i^{\text{vec}}$;
16:    Append $x_i^{\text{vec}}$ into $X^d$;
17: **end for**
18: $Z^d = F(X^d)$;  /* Transformation by the NF */
19: **for** $z_i^{\text{vec}} \in Z^d$ **do**
20:    Let $z_i$ be the sum of $z_i^{\text{vec}}$;  /* Decoder, merge features for index */
21:    Add $z_i$ into $Z$;
22: **end for**
23: **return** $Z$ for the index;

Table 2: Average transformation latency of each key with different NFs. "H" and "L" correspond to the hidden dimension and the number of layers, respectively. The figure in brackets (e.g., "(12)") represents the amount of parameters of NF. All latency is measured in nanosecond (ns).

| Batch Size | 2H2L (8) | 2H4L (16) | 4H3L (32) | 4H4L (48) |
|---|---|---|---|---|
| 1 | 169.53 | 384.84 | 320.15 | 463.81 |
| 8 | 40.60 | 83.05 | 77.52 | 113.31 |
| 32 | 15.28 | 34.75 | 33.80 | 49.40 |
| 128 | 9.52 | 24.00 | 24.91 | 36.93 |
| 256 | 8.38 | 21.66 | 23.52 | 35.07 |
| 1024 | 7.40 | 19.81 | 22.21 | 33.13 |
| 2048 | 7.29 | 19.63 | 22.00 | 32.73 |

and it takes time to train (about 38 seconds). However, since normalizing flows have much better generalization, they don't need to re-train during every bulk loading phase or adjustments in the learned indexes. Therefore, the training stage can be processed offline with neural network accelerators (e.g., GPU or FPGA) when the index bulk loads for the first time or when the distribution significantly shifts. Meanwhile, the training of normalizing flows can be easily finished in the background without affecting the learned indexes.

However, the inference of normalizing flows must be finished online. The transformed keys cannot be stored, as it would cause another indexing problem. By comprehensive tests, there are a few key factors related to the inference efficiency: 1) the input dimensions 2) the number of layers 3) the hidden dimensions 4) the implementation platform. For the first three factors, we slightly search for the parameters without increasing the conflict degrees and with a low search cost. As Table 2 shows, the search space is small limited by the unacceptable transformation overheads of larger NFs. For the acceleration platform, we implement the key computations in C++ with Intel Math Kernel Library [25]. The computations in inference can be simplified as several matrix computations and nonlinear function computations.

Moreover, since keys in some datasets are already near-uniform distributed, it is unnecessary to spend extra time and memory to transform them. Therefore, the NFL uses a switching mechanism, referring to the tail conflict degree, to determine whether to use the Numerical NF for distribution transformation. Specifically, the NFL first tries to transform the input keys, and computes the tail conflict degree based on the input keys and the transformed keys, respectively. If the latter tail conflict is larger, the NFL determines not to use the Numerical NF.

### 3.3 AFLI: After-Flow Learned Index

Although different datasets have different original conflict degrees, after the transformation of normalizing flows, the conflict degrees can be kept around a low value (e.g., around 4 for the tail conflict degree). Further reduction on conflict degrees requires more efforts on the specific design of normalizing flows, which does not conform with our original goal. Meanwhile, existing learning indexes set some empirical hyper-parameters (e.g., the maximum amount of keys assigned to a node) in order to guarantee an acceptable performance on data sets with a high conflict degree at the cost of the performance degradation on data sets with a low conflict degree. Therefore, those points motivate us to propose a robust after-flow learned index, called AFLI, that takes the characteristics of transformed data sets into considerations.

The main idea of AFLI is to buffer local conflicts. For small conflicts at some positions, the conflicted keys are stored in a bucket. The linear model directly built to handle the conflicted keys lacks generalization capability [42], as there are not enough keys to provide information. However, too many keys stored in the buckets would degrade the query performance of AFLI. Therefore, AFLI needs to make a trade-off between the maximum number of keys stored in the bucket and the generalization capability of the linear model.

*3.3.1 Structure of AFLI.* The overall structure of AFLI is shown in Figure 5. There are three types of nodes, the model node, the bucket, and the dense node.

**Model Node:** The model node consists of an array of entries and a linear model. There are four types of entries,

- *Empty Slot.* The entry is an unused slot, waiting to be filled by a pair of key-payload or a pointer.
- *Data Slot.* The entry contains a pair of key-payload.
- *Bucket Pointer.* The entry is a bucket pointer that stores the address of a bucket.
- *Node Pointer.* The entry is a node pointer that stores the address of a model node or a dense node. Especially, there might be some node pointers that contain the same address, which is called the duplicated node pointer. Since some keys are mapped to different but adjacent positions, for better generalization capability of the linear model, we assign them to the same model child node, so that corresponding pointers all point to the same address.

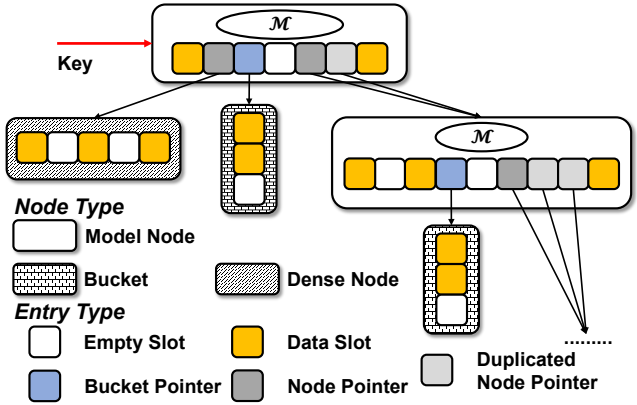

Figure 5: Structure of AFLI.

To make linear models produce fewer conflicts, we also scale the positions used to train linear models according to the scaling relationship between keys.

**Bucket:** The bucket only consists of a short data array. The maximum size of the data array is determined by the tail conflict degree $D_{\mathcal{M}}^{Y}$, but will be kept within a preset threshold range. We provide two kinds of buckets, the linear bucket (default) and the ordered bucket.

**Dense Node:** The dense node also only consists of a data array. Different from the bucket, the dense node is generally a bit larger than the bucket, but much smaller than the model node. The data array is an ordered and gapped array, the maximum number of gaps is set to $D_{\mathcal{M}}^{Y}$. Although in the Figure 5, the gap is represented by the empty space, we actually fill the slot with the data that is the closest to the front of the gap. With such fillings, we do not need an additional bitmap to indicate whether a position is empty or not, but directly compare adjacent positions to check whether store the same data.

**Analysis:** Most of nodes in the index will be model nodes. Data stored in the model nodes are directly placed in the predicted positions, which guarantees that the predictions are all precise in the model nodes. If the key space as a whole satisfies the near-uniform distribution, but with many localities which means that keys in small sub key space are too close to be distinguished by a linear model, the index will create many buckets to handle local conflicts. When the index fails to build a model node as all keys in the node are too close (i.e., the slope of the fitted linear model is 0), the index does not further partition the key space but allocates a dense node to store them. The insight of designing buckets or dense nodes is to buffer keys without increasing the height of the index, until they have enough difference to distinguish (dense nodes) or they can provide enough information to build a linear model with better generalization capability (buckets).

*3.3.2 Operations of AFLI.* In this section, we describe how our index performs lookups, insertions, and the algorithm to rebuild a bucket or a dense node into a model node.

**Queries:** To look up a key, the index recursively traverses the tree until the result is found. Firstly, the index feeds the queried key into the root node, which can only be a model node or a dense node, as the bucket is always extended from the model node. (1) If the node is a model node, the queried key will be first fed into the linear model. Based on the prediction of the linear model, we check the type of entry in the position based on two bitmaps. If it is an empty slot, the index doesn't contain the key. If it is a data slot, we compare the stored key with the queried key. If it is a bucket pointer, we search for the key further in the bucket. If it is a node pointer, we do the same procedure in the child node. (2) If the node is a dense node, we use the binary search to find the result.

It is important to note that the predicted positions in model nodes are all precise positions, which means that there is no extra local search in model nodes, so that the query performance can be significantly improved. For queries in buckets, the linear search is enough to provide comparable efficiency. When looking up in dense nodes, we directly use the binary search for efficiency.

**Insertions:** For inserting a pair of key and payload, starting from the root node, the index performs a recursive procedure based on the node type,

- If the key-payload pair is inputted into a model node, the linear model first predicts a position based on the input key. Considering the type of the entry in the predicted position, (1) If the entry is an empty slot, we directly store the key-payload pair in the position and modify corresponding bitmaps. (2) If the entry is a data slot, which means there is already a key-payload pair, so we create a bucket to store these two conflicted key-payload pairs, set the entry to the bucket pointer, and modify corresponding bitmaps. (3) If the entry is a bucket pointer or a node pointer, we insert the key-payload pair into the bucket or the child node.
- For inserting the key-payload pair into a bucket, the key-payload pair will be directly appended to the tail of the stored data. If the bucket is in the ordered mode, the insertion will be performed as the insertion sorts.
- To insert the key-payload pair into a dense node, the index first performs a binary search on the array. If the entry in the position is an empty slot, we directly insert the key-payload pair. If the entry is a data slot, we shift the data to the closest empty slot, then insert the key-payload pair.

When the bucket or the dense node has no empty slots, we try to convert it to a model node by a **Modeling** operation (more details in the next section).

**Modelling the Bucket or the Dense Node:** As shown in the Figure 6, although there are two types of nodes that need to be modelled as a model node, the modelling process are the same except for sorting the bucket first, as the inputs can be formulated as an array of key-payload pairs. Therefore, given an array of key-payload pairs, we use Algorithm 3.2 to implement the modelling operation.

In Algorithm 3.2, we first try to build a linear model using the *Linear Regression* (Line 1). Then, we determine the node type based on the prediction results. If the slope of the linear model is 0 or all keys are mapped to the same position, we build a dense node for them without further partition on the key space (Line 2-4). The reason is that those keys are too close so that the rounding operation rounds them to the same integer. If we successfully build a linear model, we build a model node for all keys. We compute the conflict degrees $\mathfrak{D}$ of each predicted position (Line 6), and use it to determine whether store the key in the data slot, or build a bucket, or build a batch of keys in child nodes. For more precise placement, the size of the entry array is the minimum value between the $\alpha$ times the number of keys and the predicted size (Line 7). Then, we iterate each predicted position, determining the entry type of each position. If the conflict degree of a position is 1, we directly store the key in the data slot (Line 10-13). If the conflict degree is larger than 1 but smaller than the maximum size of the bucket $D^Y_{\mathcal{M}_L}$, we store the conflicted keys within the position into a bucket (Line 14-17). Otherwise, for the case where the number of conflicted keys is larger than $D^Y_{\mathcal{M}_L}$, we collect the subsequent consecutive positions whose conflict degree is also larger than $D^Y_{\mathcal{M}_L}$, and allocate a new node to handle them (Line 18-21). After building the node, we duplicate the node pointer and assign the address to other collected positions (Line 22). The modelling operation will be recursively performed.

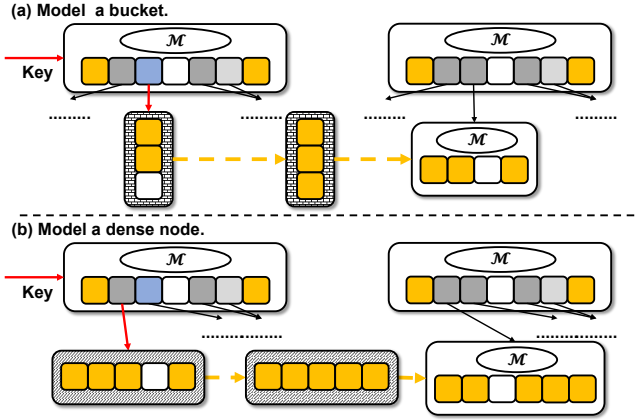

Figure 6: Modelling a bucket or a dense node.

**More Operations**:

- *BulkLoad.* The bulkload operation first computes the tail conflict degree $D^Y_{\mathcal{M}_L}$, then follows the same procedure as the modelling operation in Algorithm 3.2. The returned result is the root node.
- *Update.* An update for the payload can be finished by a lookup and an in-place update.
- *Delete.* The deletion can be implemented by a lookup on the deleted key and a modification on the corresponding node. The modification on the model node is to unset the corresponding bit in the bitmap. The modification on the bucket or the dense node is to overwrite the deleted key with the following keys.

---

**Algorithm 3.2** Modelling($\{\langle x_1, v_1 \rangle, \ldots, \langle x_n, v_n \rangle\}, \mathfrak{n}$)

**Input:** An array of sorted key-payload pairs $\{\langle x_1, v_1 \rangle, \ldots, \langle x_n, v_n \rangle\}$, the node pointer $\mathfrak{n}$ that points to the node storing all key-payload pairs, the space amplification factor $\alpha$.

**Output:** The node pointer $\mathfrak{n}$ of the model node.

1: Build a linear model $\mathcal{M}_\mathfrak{n}$ using the input keys and scaled positions.
2: **if** $\mathcal{M}_\mathfrak{n}.a == 0$ or all keys are mapped to the same position **then**
3:     $\mathfrak{n}.size = n + D^Y_{\mathcal{M}_L}$;
4:     Allocate an array of size $\mathfrak{n}.size$ to $\mathfrak{n}.\mathcal{E}$, and insert all key-payload pairs, evenly gapped by a total of $D^Y_{\mathcal{M}_L}$ gaps;
5: **else**
6:     Compute the conflict degrees $D$ of each predicted position;
7:     $\mathfrak{n}.size = \text{MIN}(\lfloor n \cdot \alpha \rfloor, \text{pos}_{last} - \text{pos}_{first} + 1)$;
8:     Allocate an array of size $\mathfrak{n}.size$ to $\mathfrak{n}.\mathcal{E}$;
9:     $i = 0$;
10:     **for** pos $\in$ all predicted positions **do**
11:         **if** $D[\text{pos}] == 1$ **then**
12:             $\mathfrak{n}.\mathcal{E}[\text{pos}] = \langle x_i, v_i \rangle$;
13:             $i = i + 1$;
14:         **else if** $D[\text{pos}] < D^Y_{\mathcal{M}_L}$ **then**
15:             Build a bucket $\mathfrak{b}$ of maximum size $D^Y_{\mathcal{M}_L}$, storing key-payload pairs $\{\langle x_i, v_i \rangle, \ldots, \langle x_{i+D[\text{pos}]}, v_{i+D[\text{pos}]} \rangle\}$;
16:             $\mathfrak{n}.\mathcal{E}[\text{pos}] = \mathfrak{b}$;
17:             $i = i + D[\text{pos}]$;
18:         **else if** $D[\text{pos}] >= D^Y_{\mathcal{M}_L}$ **then**
19:             Iterate the subsequent positions $\text{pos}_{seq}$, where $D[\text{pos}_{seq}] > D^Y_{\mathcal{M}_L}$, and sum all conflict degrees $\mathfrak{D}_{seq}$;
20:             Allocate a new node to $\mathfrak{n}.\mathcal{E}[\text{pos}]$;
21:             **Modelling**($\{\langle x_i, v_i \rangle, \ldots, \langle x_{i+\mathfrak{D}_{seq}}, v_{i+\mathfrak{D}_{seq}} \rangle\}, \mathfrak{n}.\mathcal{E}[\text{pos}]$);
22:             For all positions from pos + 1 to $\text{pos}_{seq}$, set the pointer value of each position to $\mathfrak{n}.\mathcal{E}[\text{pos}]$;
23:             $i = i + \mathfrak{D}_{seq}$;
24:         **end if**
25:     **end for**
26: **end if**
27: **return** $\mathfrak{n}$

---

## 4 EVALUATION

### 4.1 Experimental Setup

*4.1.1 Datasets and Workloads.* We choose seven representative datasets used in [6, 20, 42] to evaluate the effectiveness of the proposed NFL and index. For simplicity, all data sets consist of about 200 million unique keys. The key type is 'double', and the payload type is 'int64'. The detailed information is as follows, (1) The *longitudes* (LTD) dataset consists of the longitudes of locations around the world from Open Street Maps [31]. (2) The *longlat* (LLT) dataset consists of compound keys that combine longitudes and latitudes from Open Street Maps by applying the transformation $k = 180 \cdot \text{FLOOR}(longitude) + latitude$ to every pair of longitude and latitude. (3) The *lognormal* (LGN) dataset is a synthetic dataset whose keys are sampled from a lognormal distribution with $\mu = 0$ and $\sigma = 2$, multiplied by $10^9$ and rounded down to the nearest integer. (4) The *YCSB* dataset is generated from the YCSB benchmark [4], whose keys representing user IDs. (5) The *amazon* (AMZN) dataset consists of book sale popularity data on the Amazon [35]. (6) The *facebook* (FB) dataset is an upsampled version of a Facebook

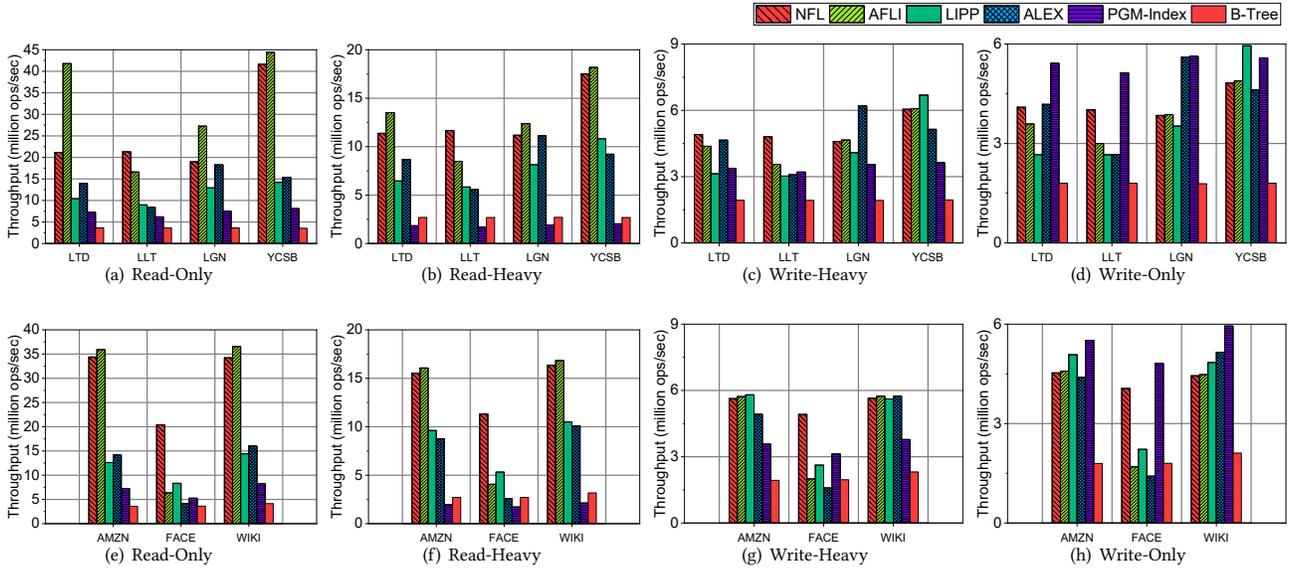

Figure 7: Throughput of NFL and Baselines.

user ID dataset [36]. (7) The *wikipedia* (WIKI) dataset is Wikipedia article edit timestamps [10].

We construct four types of workloads based on each kind of dataset. All workloads include two phases, the loading phase and the running phase. For the loading phase, we use the bulk loading operation to load 50% key-payload pairs of a dataset. For the running phase, we generate requests based on different operation ratios. (1) The *read-only* workloads only consist of query operations. (2) The *read-heavy* workloads consist of 80% query operations and 20% insertion operations. (3) The *write-heavy* workloads consist of 20% query operations and 80% insertion operations. (4) The *write-only* workloads only consist of insertion operations. For all workloads, the requested key is sampled from the given dataset based on a Zipfian distribution, and all inserted keys are in the key space consisting of all bulk-loaded keys, which means all insertions are known-key-space insertions. We run each workload for 5 times, and collect the averaged *Throughput, Tail Latency, Index Size*.

*4.1.2 Baselines.* We compare our NFL with existing state-of-the-art indexes, (1) LIPP [42], an updatable learned index with precise positions, which creates new nodes to handle conflicted keys. (2) ALEX [6], an in-memory, updatable learned index, which uses gaps to handle new insertions. (3) PGM-Index [12], a fully-dynamic compressed learnded index with provable worst-case bounds. (4) B-Tree [26], an efficient B-Tree implemented by Google. These source codes are publicly available, and we download their codes and evaluate them with their default hyper-parameters.

*4.1.3 Environment and Parameters.* We implement our proposed NFL and AFLI in C++, and perform the inference of NF by Intel Math Kernel Library [25]. We perform our evaluation via single-thread, and compile all source codes with GCC 9.3.0 in O3 optimization mode. All experiments are conducted on an Ubuntu 20.04 Linux machine with a 2.9 GHz Intel Core i7-10700 (8 cores) CPU and 64GB memory.

For NF, we train a modified B-NAF in PyTorch. The B-NAF is set to two layers, two input dimensions, two hidden dimensions, and a normal distribution with the variance of $10^{16}$ as the latent distribution. The training step is performed on the NVIDIA GeForce RTX 3080 with 10 GB GPU memory and 64 GB main memory. We only sample 10% bulk-loaded keys for three times to train the NF. We set the batch size to 256. For the proposed learned index, we limit the maximum size of buckets to no more than 6.

### 4.2 Throughput

Figure 7 shows the average throughput of NFL and other baselines on different workloads. To show the effectiveness of NF, we also evaluate our proposed AFLI. Since transforming the workloads with small conflict degrees (i.e., YCSB, AMZN, WIKI) increases the tail conflict degree, our NFL determines not to use NF.

**Read-Only Workloads.** The performance on read-only workloads reflects the approximation quality during the loading phase. Figure 7 (a) and (e) shows that our proposed NFL can achieve 2.34x, 2.46x, 3.82x, and 7.45x improvements on the throughput on average compared to LIPP, ALEX, PGM-Index, and B-Tree, respectively. Especially, for workloads with large conflict degrees (i.e., LLT and FB), our proposed NFL can achieve up to 2.41x, 3.70x higher throughput than LIPP, and ALEX, respectively. Such improvements benefit from both the Numerical NF and the proposed AFLI. The NF transforms the keys into another key space, where keys are more near-uniform distributed, forming a more linear CDF curve. Therefore, based on the transformed key space, the learned index can use less linear models to make a better approximation, and the hierarchical structure is much lower, leading to a significant improvement on throughput.

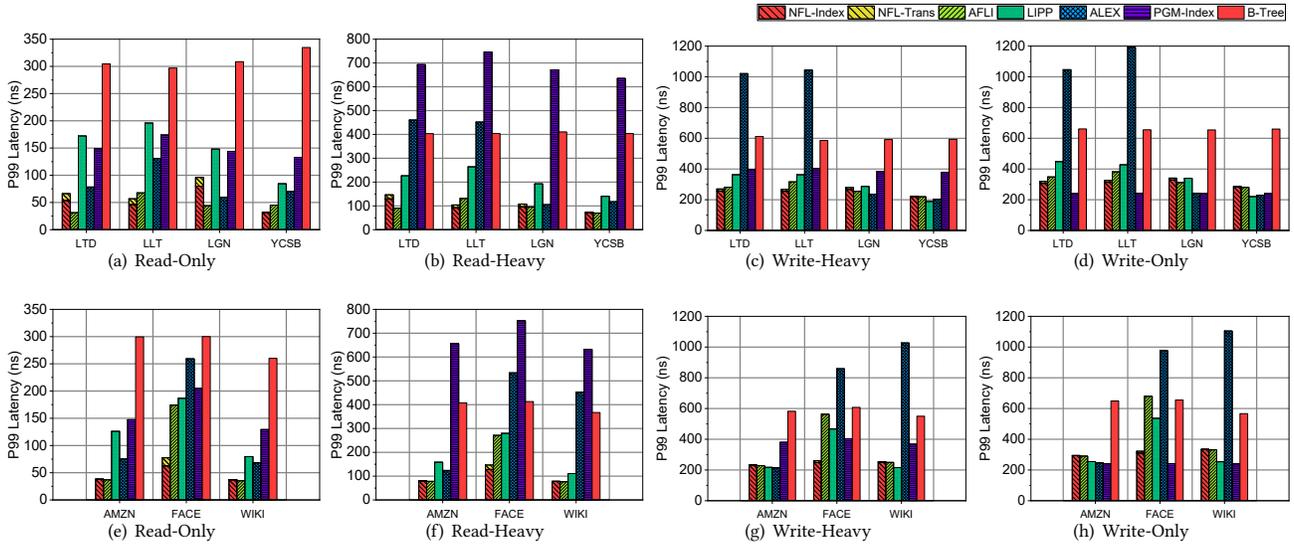

Figure 8: Tail Latency (P99) of NFL and Baselines.

Compared to the experimental results of the proposed AFLI on workloads with large conflict degrees, the NF can revisionachieve 2.25x improvement on the throughput. However, on workloads with a little reduction on the tail conflict degree (e.g., LTD, LGN), NF degrades the throughput. This is due to that compared to the improvements of the performance on transformed keys, the online inference of NF introduces a non-negligible overhead. On YCSB, AMZN, and WIKI, the NFL disables the NF due to that the distribution transformation does not reduce the tail conflict degree, so the NFL achieves almost the same performance as the proposed AFLI.

**Read-Write Workloads.** The performance on the read-write workloads (i.e., read-heavy and write-heavy workloads) shows the adjustment efficiency of the learned index during the loading phase. Figure 7 (b) and (f) shows that NFL can still achieve the same improvements on the throughput by 72.22%, 101.05%, 611.48%, and 389.45% on average compared to LIPP, ALEX, PGM-Index, and B-Tree, respectively. Although the improvements on write-heavy workloads degrade, NFL can still improve the throughput by 29.10%, 39.28%, 50.88% and 162.92% on average compared to LIPP, ALEX, PGM-Index, and B-Tree. The reason for the drop on the improvements is that more keys are buffered on the bucket, and most of the buckets are almost full, resulting in a decrease in linear search performance on the bucket. However, compared to other learned indexes which perform lots of complex and expensive adjustments (reflected in Figure 8), NFL uses tiny buckets to buffer keys and later builds linear models for the buckets which is a quite cheap adjustments.

Different with the high throughput on the read-only workloads of LTD and LGN, proposed AFLI's performance on the read-write workloads of LTD and LGN is comparable to NFL. The reason lies in that the former performance of proposed AFLI is like the case of overfitting in Machine Learning, which means that almost all linear models are well fitted and there are fewer empty slots prepared for new insertions, thus new insertions are more likely inserted into the buckets. However, since the NF have much better generalization capability than linear models, the data array of each node will be larger than before, NFL can produce less conflict degrees and construct a lower hierarchical index, which maintains the comparable performance.

**Write-Only Workloads.** Figure 7 (d) and (h) show that except for 21.43% degradation compared to PGM-Index, our proposed NFL improves the throughput by 22.65%, 28.30%, and 131.58% compared to LIPP, ALEX, and B-Tree, respectively. The high insertion performance of PGM-Index benefits from the LSM-Tree structure, where a small buffer of size 128 is used to receive new insertions. However, PGM-Index needs to perform the compaction operations periodically which is a quite expensive operation. The long tail latency in Figure 8 (d) and (h) also proves this point. Since the tail conflict degree barely changes after the running phase (see Table 3 below), the tiny buffer in NFL is enough to absorb locally conflicted keys without introducing more internal adjustments, thus NFL can still maintain high throughput after insertions. Moreover, with the transformation of NF, all keys in the key space become more uniform, leading to the reduction on the tail conflict degree (e.g., LTD, LGN), thus more keys can are buffered in the buckets and the performance of NFL can be further improved.

### 4.3 Tail Latency

To further display the impact of internal adjustments, we also evaluate our proposed NFL and other baselines on the 99% (P99) latency. In each run, we collect the latency of each batch of operations, sort the latencies in the ascending order, and report the 99-th percentile batch latency divided by the batch size in Figure 8. We also collect the 99.99% (P99.99) latency and the maximum (max) latency on some representative workloads.

**Read-Only Workloads.** The tail latency on the read-only workloads indicates the worst case of querying the learned index, corresponding to the approximation quality of the most conflicted sub key space. Figure 8 (a) and (e) shows that NFL can reduce the P99 latency by 58.68%, 32.89%, 62.73%, and 80.77% compared to LIPP, ALEX, PGM-Index, and B-Tree, respectively. Meanwhile, on the workloads with large conflict degrees (e.g., LLT, FB), NFL can still keep the P99 latency under 80 ns, while others produce over 100ns P99 latency due to expensive internal adjustments. The main reason for short P99 latency is that the normalizing flow can alleviate the worst case, reducing the conflict differences between positions and avoiding a large amount of conflicts in certain positions. The long P99 latency of proposed AFLI on the workloads LLT and FB also prove that without the transformation of NF, the learned index can be an unbalanced tree due to locally large conflicts, thus querying on those deep leaf nodes in the tree will introduce long latency.

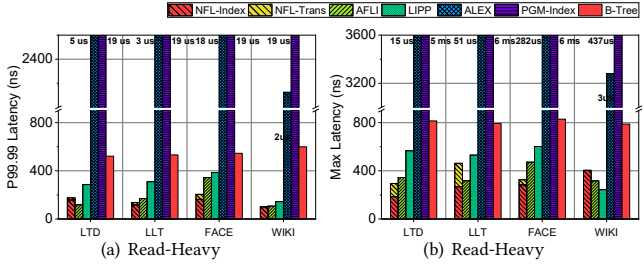

**Figure 9: P99.99 latency and max latency of NFL and Baselines.**

**Table 3: Tail conflict degrees of workloads. The suffix "(L)" and "(R)" represent the results after the loading phase and running phase, respectively. Rows in orange are results after the transformation of NF.**

|          | LTD | LLT | LGN | YCSB | AMZN | FB  | WIKI |
|----------|-----|-----|-----|------|------|-----|------|
| Tail (L) | 8   | 146 | 14  | 3    | 4    | 386 | 2    |
| Tail (R) | 7   | 147 | 13  | 3    | 4    | 454 | 1    |
| Tail (L) | 4   | 4   | 4   | 4    | 4    | 4   | 4    |
| Tail (R) | 4   | 5   | 4   | 4    | 4    | 4   | 4    |

**Read-Write Workloads.** For read-write workloads (Figure 8 (b), (c), (f), and (g)), NFL can reduce the P99 latency by 26.64%, 45.05%, 59.49%, and 65.31% compared to LIPP, ALEX, PGM-Index, and B-Tree, respectively. The long tail latency on such read-write workloads is mainly caused by two aspects, complex internal adjustments and querying on deep leaf nodes. The latter is the same as the case on read-only workloads. As the amount of insertions increases, the cost and frequency of internal adjustments for our proposed learned indexes, LIPP, and ALEX also increases, thus resulting in long P99 latency. The reduction on P99 latency for PGM-Index from read-heavy workloads to write-heavy workloads is due to the poor query performance and the high insertion performance of the LSM-Tree structure used in PGM-Index.

**Write-Only Workloads.** Figure 8 (d) and (h) show that NFL can reduce the P99 latency by 2.26%, 27.92%, and 50.48% on write-only workloads compared to LIPP, ALEX, and B-Tree, respectively. Due to the LSM-Tree's high writing performance, NFL produce 32.09% higher P99 latency than that of PGM-Index. ALEX performs expensive internal adjustments (e.g., merging or splitting nodes). Meanwhile, the cost of the shifting operations in ALEX will increase as the number of keys in a ALEX's node increases. LIPP produces comparable P99 latency to our proposed NFL, since both indexes adopt the precise placement without shifting. The difference between LIPP and NFL is that LIPP directly builds a new node for conflicted keys while NFL uses a bucket to buffer the conflicted keys first. This building operations is a little expensive compared to appending to a data array, thus NFL can produce a bit lower P99 latency than LIPP.

**P99.99 Latency and Max Latency.** We also collect the P99.99 latency and the max latency in one run. Due to the space limitation, we only present the results on the representative workloads. Figure 9 (a) and (b) shows that our proposed NFL can produce the lowest latency both on P99.99 latency and max latency, which also proves the effectiveness of the NF.

### 4.4 Other Results

*4.4.1 Conflict Degree.* The conflict degree reflects the transformation quality of NF. Table 3 shows that with the transformation of NF, the conflict degree can be significantly reduced. Although NF only uses 10% of bulk-loaded keys, after inserting around 100 million new keys, the tail conflict degree is around 4. For datasets with low conflict degrees (e.g., YCSB, WIKI), the NF increases the conflict degrees. This is due to the limitation of NF, which means that in our setting, the transformation quality of the NF has almost reached to its upper limit. But if we can design a better NF, the conflict degrees can be further reduced.

*4.4.2 Bulk Loading Time.* Figure 10 demonstrates the time cost of loading phase. For our proposed NFL, the time cost includes the time cost of the online transformation of bulk-loaded keys and the time cost of the bulk loading of the index. Our proposed AFLI achieves the fastest bulk loading compared to other baselines except for PGM-Index. Although NFL takes 77% time to transform keys and requires 2.25x, 0.86x, 2.81x time to bulk load compared to LIPP, ALEX, and B-Tree, NFL can make a better approximation of the CDF leading to a faster loading on the learned index. Figure 7 also proves that the built learned index can deliver high performance especially for highly non-linear workloads. Compared to PGM-Index, the bulk-loading time cost of NFL is much larger. However, the overall performance of NFL is much higher than PGM-Index. In addition, the average time cost of NFL to bulk load 100 million keys is 13.42 seconds, which is acceptable and worthwhile considering the high performance.

*4.4.3 Index Size.* Figure 11 shows the normalized final index size of existing learned indexes after the running phase of write-heavy workloads. Since the great improvements of the state-of-the-art learned indexes benefit from precise placements which require lots of gaps [6] in the data array, we evaluate the overall index size, including the sum size of allocated gaps, rather than the model size in the learned index. The index size of NFL is 2.26x, 3.1x times than the index size of ALEX and PGM-Index, respectively. However, the index size of NFL is only 0.51 of LIPP's index size. Both NFL and LIPP create more space to support precise placements, but NFL

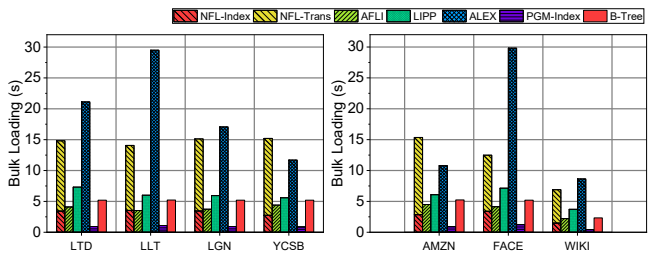

Figure 10: Bulk loading of NFL and Baselines.

makes a better trade-off between allocating more space and fitting linear models better.

## 5 RELATED WORK

### 5.1 Learned index

Recent work on the Learned Index [22] brings a new direction to the indexing field, which leverage machine learning models to predict the mapping between keys and positions. Inspired by [22], a series works fully explore the potential of machine learning models on the indexing problems. Both FITing-Tree [13] and PGM-Index [12] approximate the CDF using piece-wise linear functions to restrict the prediction error in a given bound. Without a given error bound, ALEX [6] places data in the predicted position of a gapped array, which makes a better approximation of the CDF. To support more precise placements, LIPP [42] defines the conflict degree to reflect the approximation quality and directly create new nodes for the conflicted keys. LIPP [42] achieves the state-of-the-art performance compared to previous learned indexes. However, our work indicates that the performance of learned indexes can be further improved by applying a distribution transformation on the input keys, which fundamentally improve the approximation quality of the CDF. There are several works focusing on improving the approximation by Gap Insertion [24], Log-Error Regression [11]. They don't consider the impact of different key distributions. Other works investigate the effectiveness of learned indexes on multi-dimensions or spatial queries [7, 23, 30, 39], construction efficiency [21], string index [40], multi-threads [38]. They are different from our scope. Other works also combine the learned index with modern applications [5, 37], new hardware [3, 27]. Our works can also be adapted to their scope.

### 5.2 Normalizing Flows

Normalizing flows are popularised in the context of variational inference [34] and density estimation [8]. The natural and most obvious use of normalizing flows is to perform density estimation, which is the same scope of the problem in learned indexes. The key point of normalizing flows is how to design the generators, in terms of invertible transformation and efficient computations. Dinh et al. introduced a *coupling* method to improve the expressiveness of the transformation [8]. The key of [8] is how to partition $x$, which inspires a series of work [9, 17]. However, such a coupling method requires the dimension of input keys to be at least greater than 2, which is not suitable in our problem. Another widely used flow structure is auto-regressive flow, first proposed by Kingma et al. in IAF [18]. Huang et al. replace the affine univariate transformations of MAF [32] and IAF [18] with a more general class of invertible univariate transformations called monotonic neural networks, and achieve a better performance. B-NAF [2] improves the structure of NAF by using a single feed-forward network to model the bijections, and becomes much more compact than NAF while remaining comparable universal approximation. For the first time, we apply the state-of-the-art auto-regressive flows to improve the approximation of learned indexes.

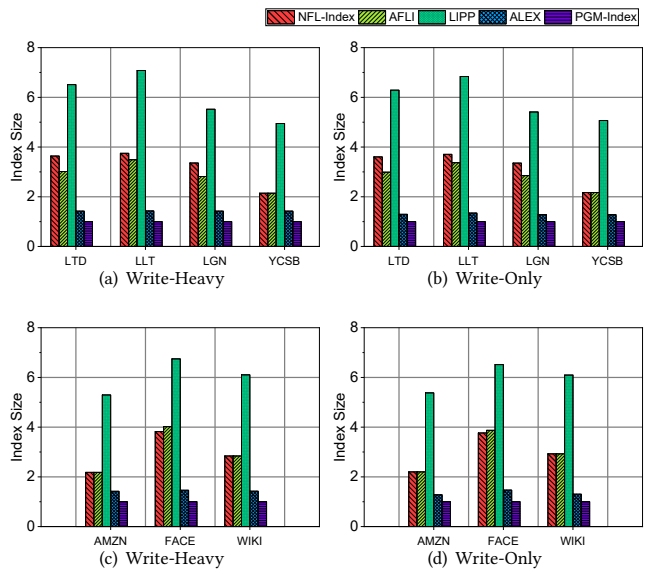

Figure 11: Index size of NFL and Baselines.

## 6 CONCLUSIONS

In this paper, we address the approximation problem by applying a distribution transformation. We present a normalizing-flow-learned index framework to transform the key space for better approximation on the CDF. We also introduce a 1D-friendly normalizing flow to achieve the distribution transformer. We further propose a lightweight and precise learned index to support efficient indexing operations. Experimental results on representative workloads demonstrate the effectiveness of our proposed framework and our proposed index.

For the future work, we plan to further investigate the potential of using normalizing flows, including co-design on normalizing flows and learned indexes, using normalizing flows on multi-dimension indexing and so on. Besides, it is also worthwhile to further accelerate the inference of normalizing flows.

# REFERENCES

[1] Jie An, Siyu Huang, Yibing Song, Dejing Dou, Wei Liu, and Jiebo Luo. 2021. ArtFlow: Unbiased Image Style Transfer via Reversible Neural Flows. In *IEEE Conference on Computer Vision and Pattern Recognition (CVPR)*. 862–871.

[2] Nicola De Cao, Wilker Aziz, and Ivan Titov. 2019. Block Neural Autoregressive Flow. In *Proceedings of the Thirty-Fifth Conference on Uncertainty in Artificial Intelligence (UAI)*, Vol. 115. 1263–1273.

[3] Leying Chen and Shimin Chen. 2021. How Does Updatable Learned Index Perform on Non-Volatile Main Memory?. In *Proceedings of the 37th IEEE International Conference on Data Engineering Workshops (ICDEW)*. 66–71.

[4] Brian F. Cooper, Adam Silberstein, Erwin Tam, Raghu Ramakrishnan, and Russell Sears. 2010. Benchmarking cloud serving systems with YCSB. In *Proceedings of the 1st ACM Symposium on Cloud Computing (SoCC)*. 143–154.

[5] Yifan Dai, Yien Xu, Aishwarya Ganesan, Ramnatthan Alagappan, Brian Kroth, Andrea C. Arpaci-Dusseau, and Remzi H. Arpaci-Dusseau. 2020. From WiscKey to Bourbon: A Learned Index for Log-Structured Merge Trees. In *Proceedings of the 14th USENIX Symposium on Operating Systems Design and Implementation (OSDI)*. 155–171.

[6] Jialin Ding, Umar Farooq Minhas, Jia Yu, Chi Wang, Jaeyoung Do, Yinan Li, Hantian Zhang, Badrish Chandramouli, Johannes Gehrke, Donald Kossmann, David B. Lomet, and Tim Kraska. 2020. ALEX: An Updatable Adaptive Learned Index. In *Proceedings of the 2020 International Conference on Management of Data (SIGMOD)*. 969–984.

[7] Jialin Ding, Vikram Nathan, Mohammad Alizadeh, and Tim Kraska. 2020. Tsunami: A Learned Multi-dimensional Index for Correlated Data and Skewed Workloads. *Proc. VLDB Endow.* 14, 2 (2020), 74–86.

[8] Laurent Dinh, David Krueger, and Yoshua Bengio. 2015. NICE: Non-linear Independent Components Estimation. In *Proceedings of the 3rd International Conference on Learning Representations (ICLR)*, Yoshua Bengio and Yann LeCun (Eds.).

[9] Laurent Dinh, Jascha Sohl-Dickstein, and Samy Bengio. 2017. Density estimation using Real NVP. In *Proceedings of the 5th International Conference on Learning Representations (ICLR)*.

[10] Wikimedia downloads. [n.d.]. http://dumps.wikimedia.org.

[11] Martin Eppert, Philipp Fent, and Thomas Neumann. 2021. A Tailored Regression for Learned Indexes: Logarithmic Error Regression. In *Proceedings of the Fourth International Workshop on Exploiting Artificial Intelligence Techniques for Data Management (aiDM)*. 9–15.

[12] Paolo Ferragina and Giorgio Vinciguerra. 2020. The PGM-index: a fully-dynamic compressed learned index with provable worst-case bounds. *Proc. VLDB Endow.* 13, 8 (2020), 1162–1175.

[13] Alex Galakatos, Michael Markovitch, Carsten Binnig, Rodrigo Fonseca, and Tim Kraska. 2019. FITing-Tree: A Data-aware Index Structure. In *Proceedings of the 2019 International Conference on Management of Data (SIGMOD)*. 1189–1206.

[14] Georgios Giannikis, Gustavo Alonso, and Donald Kossmann. 2012. SharedDB: Killing One Thousand Queries With One Stone. *Proc. VLDB Endow.* 5, 6 (2012), 526–537.

[15] Georgios Giannikis, Darko Makreshanski, Gustavo Alonso, and Donald Kossmann. 2014. Shared Workload Optimization. *Proc. VLDB Endow.* 7, 6 (2014), 429–440.

[16] Chin-Wei Huang, David Krueger, Alexandre Lacoste, and Aaron C. Courville. 2018. Neural Autoregressive Flows. In *Proceedings of the 35th International Conference on Machine Learning (ICML)*, Vol. 80. 2083–2092.

[17] Diederik P. Kingma and Prafulla Dhariwal. 2018. Glow: Generative Flow with Invertible 1x1 Convolutions. In *Advances in Neural Information Processing Systems 31: Annual Conference on Neural Information Processing Systems 2018 (NeurIPS)*. 10236–10245.

[18] Durk P. Kingma, Tim Salimans, Rafal Jozefowicz, Xi Chen, Ilya Sutskever, and Max Welling. 2016. Improved variational inference with inverse autoregressive flow. *Advances in Neural Information Processing Systems* 29 (2016), 4743–4751.

[19] Diederik P. Kingma and Max Welling. 2014. Auto-Encoding Variational Bayes. In *Proceedings of the 2nd International Conference on Learning Representations (ICLR)*.

[20] Andreas Kipf, Ryan Marcus, Alexander van Renen, Mihail Stoian, Alfons Kemper, Tim Kraska, and Thomas Neumann. 2019. SOSD: A Benchmark for Learned Indexes. *CoRR* (2019).

[21] Andreas Kipf, Ryan Marcus, Alexander van Renen, Mihail Stoian, Alfons Kemper, Tim Kraska, and Thomas Neumann. 2020. RadixSpline: a single-pass learned index. In *Proceedings of the Third International Workshop on Exploiting Artificial Intelligence Techniques for Data Management (aiDM)*. 5:1–5:5.

[22] Tim Kraska, Alex Beutel, Ed H. Chi, Jeffrey Dean, and Neoklis Polyzotis. 2018. The Case for Learned Index Structures. In *Proceedings of the 2018 International Conference on Management of Data (SIGMOD)*. 489–504.

[23] Pengfei Li, Hua Lu, Qian Zheng, Long Yang, and Gang Pan. 2020. LISA: A Learned Index Structure for Spatial Data. In *Proceedings of the 2020 International Conference on Management of Data (SIGMOD)*. 2119–2133.

[24] Yaliang Li, Daoyuan Chen, Bolin Ding, Kai Zeng, and Jingren Zhou. 2021. A Pluggable Learned Index Method via Sampling and Gap Insertion. *CoRR* (2021).

[25] Intel Math Kernel Library. [n.d.]. https://www.intel.com/content/www/us/en/developer/tools/oneapi/onemkl.html.

[26] The C++ B-Tree library implemented by Google. [n.d.]. https://code.google.com/archive/p/cpp-btree.

[27] Baotong Lu, Jialin Ding, Eric Lo, Umar Farooq Minhas, and Tianzheng Wang. 2021. APEX: A High-Performance Learned Index on Persistent Memory. *CoRR* (2021).

[28] Darko Makreshanski, Georgios Giannikis, Gustavo Alonso, and Donald Kossmann. 2018. Many-query join: efficient shared execution of relational joins on modern hardware. *VLDB J.* 27, 5 (2018), 669–692.

[29] Darko Makreshanski, Jana Giceva, Claude Barthels, and Gustavo Alonso. 2017. BatchDB: Efficient Isolated Execution of Hybrid OLTP+OLAP Workloads for Interactive Applications. In *Proceedings of the 2017 ACM International Conference on Management of Data (SIGMOD)*. 37–50.

[30] Vikram Nathan, Jialin Ding, Mohammad Alizadeh, and Tim Kraska. 2020. Learning Multi-Dimensional Indexes. In *Proceedings of the 2020 International Conference on Management of Data (SIGMOD)*. 985–1000.

[31] OpenStreetMap on Amazon AWS. [n.d.]. https://registry.opendata.aws/osm/.

[32] George Papamakarios, Iain Murray, and Theo Pavlakou. 2017. Masked Autoregressive Flow for Density Estimation. In *Advances in Neural Information Processing Systems 30: Annual Conference on Neural Information Processing Systems 2017 (NeurIPS)*. 2338–2347.

[33] Robin Rehrmann, Carsten Binnig, Alexander Böhm, Kihong Kim, Wolfgang Lehner, and Amr Rizk. 2018. OLTPShare: The Case for Sharing in OLTP Workloads. *Proc. VLDB Endow.* 11, 12 (2018), 1769–1780.

[34] Danilo Jimenez Rezende and Shakir Mohamed. 2015. Variational Inference with Normalizing Flows. In *Proceedings of the 32nd International Conference on Machine Learning (ICML)*, Francis R. Bach and David M. Blei (Eds.), Vol. 37. 1530–1538.

[35] Amazon sales rank data for print and kindle books. [n.d.]. https://www.kaggle.com/ucffool/amazon-sales-rank-data-for-print-and-kindle-books.

[36] Peter Van Sandt, Yannis Chronis, and Jignesh M. Patel. 2019. Efficiently Searching In-Memory Sorted Arrays: Revenge of the Interpolation Search?. In *Proceedings of the 2019 International Conference on Management of Data (SIGMOD)*. 36–53.

[37] Mihail Stoian, Andreas Kipf, Ryan Marcus, and Tim Kraska. 2021. PLEX: Towards Practical Learned Indexing. *CoRR* (2021).

[38] Chuzhe Tang, Youyun Wang, Zhiyuan Dong, Gansen Hu, Zhaoguo Wang, Minjie Wang, and Haibo Chen. 2020. XIndex: a scalable learned index for multicore data storage. In *Proceedings of the 25th ACM SIGPLAN Symposium on Principles and Practice of Parallel Programming (PPoPP)*. 308–320.

[39] Haixin Wang, Xiaoyi Fu, Jianliang Xu, and Hua Lu. 2019. Learned Index for Spatial Queries. In *Proceedings of the 20th IEEE International Conference on Mobile Data Management (MDM)*. 569–574.

[40] Youyun Wang, Chuzhe Tang, Zhaoguo Wang, and Haibo Chen. 2020. SIndex: a scalable learned index for string keys. In *Proceedings of the 11th ACM SIGOPS Asia-Pacific Workshop on Systems (APSys)*. 17–24.

[41] Valentin Wolf, Andreas Lugmayr, Martin Danelljan, Luc Van Gool, and Radu Timofte. 2021. DeFlow: Learning Complex Image Degradations From Unpaired Data With Conditional Flows. In *IEEE Conference on Computer Vision and Pattern Recognition (CVPR)*. 94–103.

[42] Jiacheng Wu, Yong Zhang, Shimin Chen, Yu Chen, Jin Wang, and Chunxiao Xing. 2021. Updatable Learned Index with Precise Positions. *Proc. VLDB Endow.* 14, 8 (2021), 1276–1288.